\documentclass[aps,prc,floats,floatfix,preprint,superscriptaddress,nofootinbib]{revtex4}

\usepackage{graphicx}
\usepackage[T1]{fontenc}
\usepackage[latin9]{inputenc}
\usepackage{textcomp}
\usepackage{amsmath}
\usepackage{amssymb}
\def\be{\begin{equation}}
\def\ee{\end{equation}}

\def\dthree{\Delta^{(3)}}
\begin{document}

\title{
  Pairing gaps in HFB with the Gogny D1S interaction 
}

\author{L.M.~Robledo}
\affiliation{Departamento de Fisica Teorica, Universidad Aut\'onoma de Madrid, E-28049
Madrid, Spain}
\author{R. Bernard}
\author{G.F.~Bertsch}
\affiliation{Institute for Nuclear Theory and Department of Physics,
University of Washington, Seattle, Washington, USA}

\begin{abstract}
As part of a program to study odd-$A$ nuclei in the Hartree-Fock-Bogoliubov
(HFB) theory, we have developed a new calculational tool to 
find the HFB minima of odd-A nuclei based on the gradient method and using
interactions of Gogny's form.  The HFB minimization includes
both time-even and time-odd fields in the energy functional, avoiding the 
commonly used ``filling approximation''.  Here we apply the method to 
calculate neutron pairing gaps in some representative isotope chains
of spherical and deformed nuclei, namely the $Z=8,50$ and 82 spherical 
chains and the $Z=62$ and 92 deformed chains.  We find that the 
gradient method is quite robust, permitting us to carry out systematic
surveys involving many nuclei.  We find that the time-odd field does
not have large effect on the pairing gaps calculated with the Gogny
D1S interaction.  
Typically, adding the T-odd field as a perturbation 
increases the pairing gap by $~100$ keV, but the re-minimization
brings the gap back down.  This outcome is very similar to results
reported for the Skyrme family of nuclear energy density functionals.
Comparing the calculated gaps with the experimental
ones, we find that the theoretical errors have both signs implying that
the D1S interaction has a reasonable overall strength.  However, we
find some systematic deficiencies comparing spherical and deformed chains
and comparing the lighter chains with the heavier ones.
The gaps for heavy spherical
nuclei are too high, while those for deformed nuclei tend to be
too low.  The calculated gaps of spherical nuclei show hardly any
$A$-dependence, contrary to the data.  Inclusion of the T-odd component of 
the interaction does not change these qualitative findings. 
\end{abstract}
 
\maketitle

\section{Introduction}

The Hartree-Fock-Bogoliubov theory is now very well developed for the Skyrme
and Gogny families of interactions.  However, up to now the calculational
tools for odd-A nuclei and for other HFB wave functions that are not
time-reversal invariant have not reached the point where systematic surveys
can be easily carried out.  The problems are less severe in the so-called
filling approximation, and that approximation has become very commonplace in
the literature.  Two of us \cite{ro11} have proposed a methodology based on
the gradient method that avoids most of these computational issues. The
first aim of this work is to demonstrate that the method is practical under
``field conditions'' of typical isotope chains across the chart of nuclides. 
The second aim is to assess the error in the filling approximation for an
interaction in the Gogny family that has been widely used, namely the D1S
interaction \cite{D1S}.

There is a very large literature on the application of HFB to 
odd-$A$ systems and the filling approximation.
Refs. \cite{ba74,ma75} first showed how to using blocking to calculate odd-$A$ 
nuclei in the HFB theory.  While exact blocking has been carried out
with realistic interactions and with schematic forces \cite{ma75}, the
HFB based on global energy density functionals has largely relied on
the filling approximation.  
There
are exceptions dealing with very specific examples like
the high-spin study of Ref \cite{vi97} with the Gogny force.
The filling approximation is defined as a full minimization of 
the HFB functional but neglecting spin-dependent and other time-odd
densities.  This is equivalent to neglecting the time-odd fields in 
the functional when expressed as products of fields and densities.

The filling approximation can also be portrayed as a statistical quantum system
where the blocked orbital and its time reversed partner share the same
probability \cite{pe08}.  This formulation may have advantages with respect
to further generalizations.
The approximation was used for example in setting the parameters
of the Gogny functional in Ref. \cite{de80}. More recent applications with 
a Gogny functional are in Ref. \cite{ro10a,ro10b}.  There was an early
study of time-odd fields with the Skyrme interaction \cite{pa76}, but
most of the recent work has used the filling approximation. Notably,
it was used for  
surveys of odd-$A$ nuclei with Skyrme energy functionals 
in \cite{be00,be09}.  Very recently, the effect of time-odd 
fields in the Skyrme functional has been re-examined in two 
surveys \cite{po10,sc10}.  The filling approximation has also been
used with the relativistic mean field theory \cite{ba06}, and the
role of the time-odd fields there have been examined in Refs.
\cite{ru99,af10}.

The physical quantity we calculate in this paper is the neutron pairing gap, defined for
odd-$N$ nuclei as
\be
\dthree_o(Z,N) ={1\over 2 } \left( B(Z,N-1) + B(Z,N+1) - 2 B(Z,N)  \right)
\ee
where $B(Z,N)$ is the (positive) binding energy of the nucleus. In the
BCS theory it is calculated as the BCS gap parameter.   In finite
nuclei there can be considerable rearrangement in the wave
functions from one nucleus to the next,  and the gaps should be determined from 
Eq. (1) using the calculated binding energies.  We use this definition in
the present paper.

We consider a representative sample of isotope chains, spanning the
nuclear size range from Oxygen to Uranium isotopes.  We include both
spherical and strongly deformed nuclei in the survey, permitting us to 
examine effects of the nuclear shape. 
As mentioned above, a particular focus in our survey 
is the validity
of
the filling approximation.  We consider this to be important to 
examine because the filling approximation can give rise to an unphysical
self-energy of the odd particle, as explained in the Appendix.  
Beyond that, our survey is extensive enough to uncover possible systematic
problems with the Gogny functional we employ.  Particular aspects
are the overall mass-dependence of the pairing gaps, and the possible
differences between pairing is spherical and deformed nuclei.  Both
these aspects can indicate non-pairing contributions to the gaps
\cite{du01,fr07}. 

\section{Computational Aspects}

The calculations reported below were carried out with a new code based
on the program HFBAXIAL written by one of us \cite{hfbaxial} to carry
out HFB calculations for Gogny-type interactions.  The
algorithm to find the HFB minima uses the analytic expression
for the derivative of the HFB energy function with respect to a 
generalized Thouless transformation \cite[Eq. (7.32)]{RS} of the HFB wave 
function. Other 
aspects of the HFBAXIAL that are important for the algorithm are
described in Ref. \cite{ro11}.  That reference also introduces the
generalization of the gradient method to wave functions with odd
particle numbers.   We defer details of our new code to later 
publication \cite{CPC}.  For the present purposes, the main
points on the computational side
are the definition of the basis states and the assumed block structure
of the Bogoliubov transformation matrices $UV$.  We use an oscillator
basis with equal oscillator length parameters in each direction.
The basis states are cylindrically symmetric, with orbitals labeled
by $n_z,n_r,m$ and $s_z$.  We assume a block structure that preserves
axial symmetry in the wave function and does not mix neutrons and 
protons.  Thus the blocks may be labeled by $t_z$, the nucleon isospin,
and $j_z$, the angular momentum about the $z$-axis.  In fact we also
have to include $-j_z$ in the same block as $j_z$ because these are coupled
by the anomalous HFB field.\\

The original HFBAXIAL code assumes that the wave functions are 
time-reversal invariant, and thus the HFB fields are even under
$T$, the time-reversal operator.  The $T$-odd fields added to
the new code arise from various terms in the Gogny interaction including
the spin-orbit and density-dependent
contact terms.  In addition, there is 
a T-odd field associated with the exchange Coulomb interaction 
as well with the two-body correction to the center-of-mass
kinetic energy.  It should be mentioned that there is an intrinsic
ambiguity in the T-odd field of the density-dependent interaction;
we evaluate this term taking the density to be real.  This term 
does not contribute to the 
pairing field in first order due to its assumed exchange character.
However, it does contribute to the mean field potential giving an
impact on $\dthree$.

Among the tests we made of the code, there are two that are
worthy of mention because they are very powerful in finding 
inconsistencies in the coding.  The first test is of the gradient method
itself.  As mentioned above, the gradient of the energy with respect
to the degrees of freedom in the $UV$ Bogoliubov transformation
is computed analytically. One can also monitor the gradient
numerically from the difference in energies when the $UV$ matrices are
changed by a small amount.  If these two are not equal to the expected
precision, there is a coding error that must be corrected.  The second
test is a very simple one.  The interaction energy must vanish if the
wave function is a one-particle state\footnote[1]
{ A one-particle wave function is necessarily of Hartree-Fock
form since the Bogoliubov transformation mixes particle numbers.}. 
The vanishing is not 
trivial, as it comes about by an exact cancellation of the
direct and exchange fields of the interaction.
This provides a good test of  
exchange part of the Gogny interaction, which is computed in a 
highly optimized code.

There are two purely numerical parameters in the calculation.  The
first is the number of oscillator shell $N_{sh}$ included in the basis.  We
follow the Madrid practice taking $N_{sh}$ in the range
10 to 14 depending on mass region as given in the Figure captions.
The oscillator length parameter is taken at a fixed
value $b=2.1$ fm for all nuclei.  Obviously one could due better 
by including more shells and allowing the oscillator parameter to 
vary.  For our purposes here, the differences in total energies largely
cancel out.  The reported pairing gaps are converged to within 
several tens of keV, which is certainly acceptable for this survey
of pairing trends and the validity of the filling approximation.

To find the HFB minima of odd-A nuclei, we start with the 
converged wave functions for the even-even nuclei
on either side of the target nucleus, taking 
oblate, prolate and spherical minima. A set of trial 
odd-A wave functions is generated from them by the usual procedure of 
exchanging the $U$ and $V$ components in one of the columns
of the $UV$ matrix.  Besides having odd particle number, these
wave functions can be characterized by the angular momentum $K$ 
about the symmetry axis, equal to $\pm j_z$ of the block in which
the $U$-$V$ interchange was carried out.  Applying the gradient
solver for each of the trial wave functions, a large set of local minima 
is obtained, many of which are identical.  We select the lowest of
these.  There is no guarantee that one will always find the global 
minimum with this protocol.  Still, for the spherical and strongly 
deformed nuclei that we calculate here, the possibility of other,
deeper minima is slight.

\section{Results}
\subsection{Spherical chains}

We begin our survey with nominally spherical nuclei, 
presenting results for neutron pairing gaps in the three semi-magic isotope chains, 
$Z=8$, 50, and 82.  The experimental data shown in the figures
are based on the Audi, et al. mass table Ref. \cite{audi} with 
some additions from Ref. \cite{audi2011}. 
We start with the lightest chain in our survey, the $Z=8$ (Oxygen)
isotopes with $N >8$.  These are calculated to have very large pairing 
gaps when the $d_{5/2}$ subshell is being filled and smaller gaps in the
upper $sd$ shell.  These qualitative features are the same in 
all three approximations in our study.  The shell differences 
is of course expected because the shell degeneracy at the Fermi
level should be a strong determinant of the gap.  There is an
especially strong decrease at $N=15$, corresponding to an open
$s_{1/2}$ shell.  This ``gap quenching'' is a general
feature of pairing gap systematics as noted in Ref. \cite[See Table III]{be09}.
From Fig. 1, we see experiment agrees with the theory at a qualitative
level, showing large gaps in the
$d_{5/2}$ shell and with a strong quenching at $N=15$. 
On a more quantitatively level, the 
calculated pairing gaps are somewhat to low in the $d_{5/2}$
filling region.
\begin{figure}[htb] 
\begin{center} 
\includegraphics[width=8cm]{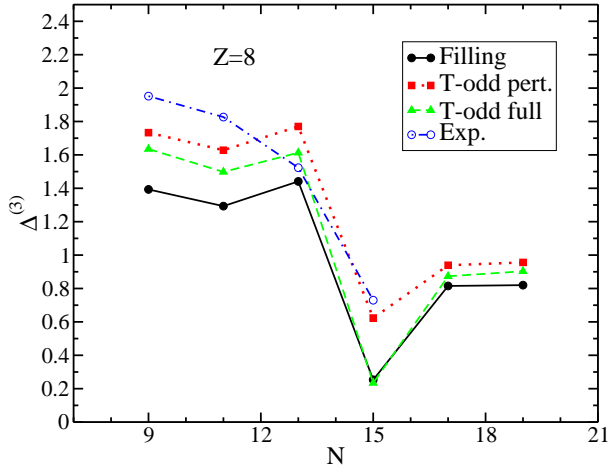} 
\caption{Neutron pairing gaps $\dthree$ in the Oxygen isotope chain.
Energies were computed in the $N_{sh} = 10 $ harmonic oscillator
space.\label{fig:oxy}} 
\end{center} 
\end{figure}

\begin{figure}[htb] 
\begin{center} 
\includegraphics[width=8cm]{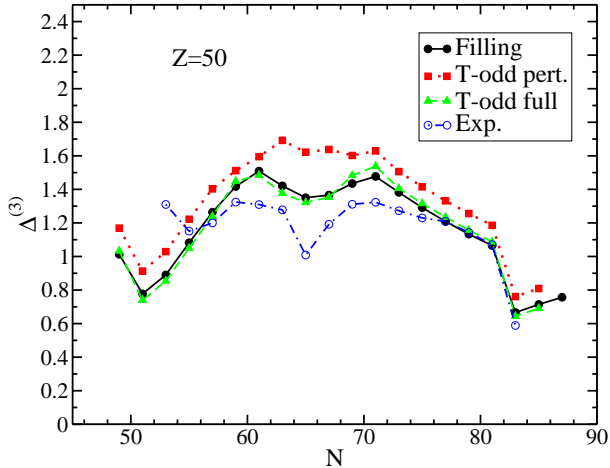} 
\caption{Neutron pairing gaps $\dthree$ in the Sn isotope chain.
Energies were computed in the $N_{sh} = 12 $ harmonic oscillator
space.
\label{fig:tin}} 
\end{center} 
\end{figure}

The next chain, the $Z=50$ Sn isotopes, has 16 measured pairing gaps and has
been the subject of many theoretical studies of pairing, eg. 
\cite{an02,yu03,ru99}. 
Our calculated gaps are shown in Fig. \ref{fig:tin}. As with the Oxygen
isotopes, the filling approximation gives very similar results to the full
calculation with $T$-odd fields. The calculated gaps start out moderate at
the beginning of the $N=50-82$ major shell, smoothly increase through the
shell with a mild dip around $N=65$.  The gaps then smoothly decrease toward the
end of the shell, and drop sharply beyond the $N=82$ shell closure.  There
is no gap quenching associated with the $s_{1/2}$ subshell, probably
because of a degeneracy with other subshells in the single-particle
spectrum. 
These features
are also present in HFB calculations based on the Skyrme \cite{be09} and the
Fayans energy functionals \cite{yu03}, so they seem to be generic for HFB with
short-ranged functionals. 
Experimentally, the gaps are rather large, and one sees the
sharp decrease at $N=83$. However, other details differ from the calculated
pattern of gaps. The predicted decrease at $N=53$ is not seen
experimentally.  The mild decrease in the middle of the shell is smoother in
the theoretical gaps than the experimental ones, which is very sharp at
$N=65$. Overall, the predicted gap is somewhat too high.  

Concerning the effect of the $T$-odd field, at a perturbative level 
it  makes
the gap even larger,  but the full calculation is
hardly distinguishable from the filling approximation.

\begin{figure}[htb] 
\begin{center} 
\includegraphics[width=8cm]{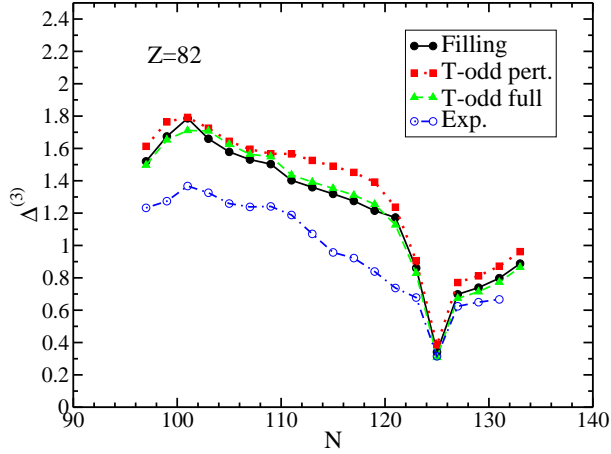} 
\caption{Neutron pairing gaps $\dthree$ in the Pb isotope chain.
Energies were computed in the $N_{sh} = 12 $ harmonic oscillator
space.\label{fig:lead}} 
\end{center} 
\end{figure}

The heaviest spherical chain in our survey is the $Z=82$ Pb isotope
chain, which also has the greatest number of measured pairing gaps
(18).  The calculated gaps are shown in Fig. \ref{fig:lead}. The theoretical 
gaps start out very large, decrease to moderate at the upper end of the 
$N=82-126$ major shell, and show the
shell quenching effect at the $p_{1/2}$ shell.  Again, all the
theory curves are similar.  Experimentally, the gaps have these
qualitative features but the overall scale for the large gaps is
markedly smaller.

In summary, the main qualitative features within an isotope chain
are reproduced by the theory with or without inclusion of the $T$-odd
field.

\subsection{Deformed nuclei}

The HFB energies of strongly deformed nuclei should be interpreted
more cautiously.    The minima now correspond
to the bandheads of the rotational bands that characterize the spectra
of these nuclei.  In a systematic study of ground state spins of
odd-$A$ nuclei \cite{bo07}, it was found that only 40\% of the
spins of deformed nuclei agreed with their self-consistent mean 
field calculations.  This raises an ambiguity in comparing
the $\dthree$ to experiment, whether to take the ground state energies
or energies of states of the same spin.  We will come back to this
point after reporting our comparison for ground state energies.

\begin{figure}[htb] 
\begin{center} 
\includegraphics[width=8cm]{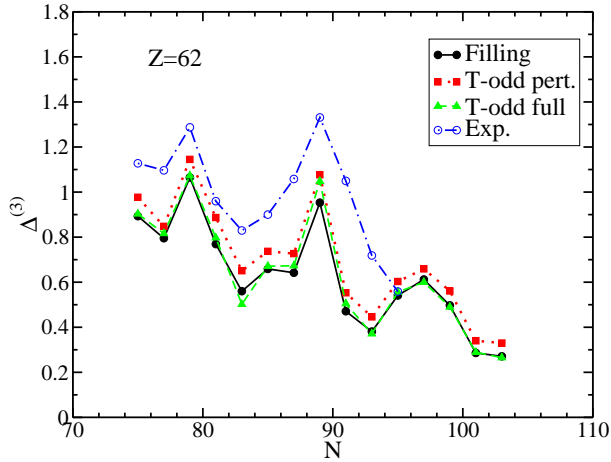} 
\caption{Neutron pairing gaps $\dthree$ in the Sm isotope chain.
Energies were computed in the $N_{sh} = 12 $ harmonic oscillator
space.\label{fig:sm}} 
\end{center} 
\end{figure}

We first show the pairing gaps for the Samarium isotope chain, well
known for showing the transition from spherical to deformed nuclei.
The calculated gaps are shown in Fig. \ref{fig:sm}.
The theoretical gaps are smaller than would be expected from 
the systematics we found for the semi-magic chains.  As in the
other cases, the $T$-odd field is significant at the perturbative
level but hardly affects the gaps in the full treatment. The experimental
data is somewhat higher than theory, but the variations along the chain follow
the theory quite well.

\begin{figure}[htb] 
\begin{center} 
\includegraphics[width=8cm]{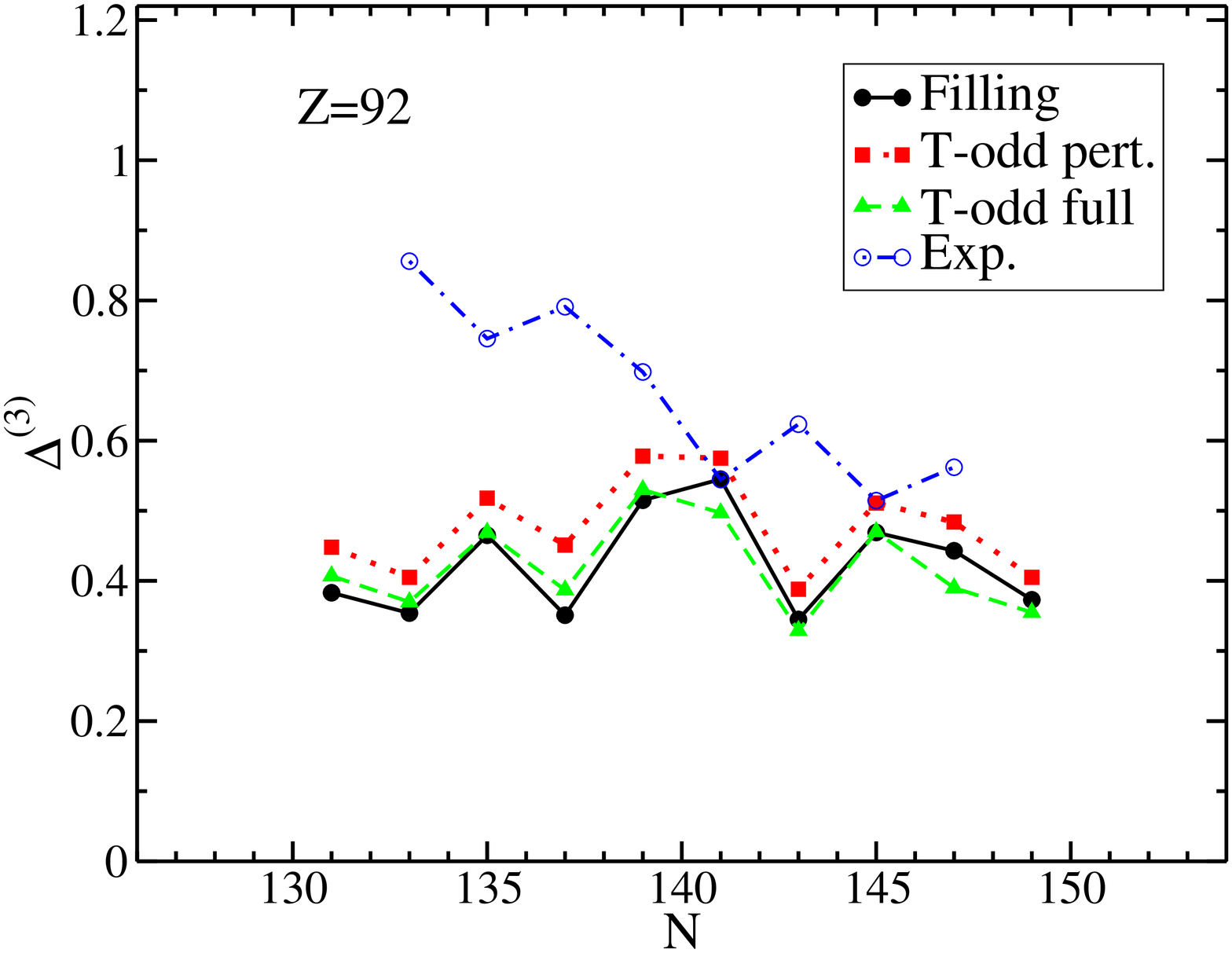} 
\caption{Neutron pairing gaps $\dthree$ in the U isotope chain.
Energies were computed in the $N_{sh} = 14 $ harmonic oscillator
space.\label{fig:u}} 
\end{center} 
\end{figure}

The last isotope chain we consider is the Uranium chain, shown in Fig.
\ref{fig:u}. The first isotope shown, $N=131$, is weakly deformed
but all the higher members have large (theoretical) quadrupole deformations.
One sees in the graph that the calculated gaps are
quite small throughout the chain.   The experimental gaps are reduced with respect to 
the systematics for the semi-magic nuclei, but not as much as 
the theory predicts.    Also, one sees that the $T$-odd field has a 
very small effect in this very heavy chain, in fact negligible on the 
scale of the accuracy of the theory.

In the two chains we treated above, 11 of the gaps were for strongly
deformed nuclei defined by the criterion that the calculated deformation
$\beta_2$ satisfies $\beta_2 > 0.2$.  Of these, the $j_z$ and parity
of the ground state agrees with experiment in five cases, including
one where there is a near degeneracy of the lowest states.  This is
roughly the proportion as found in the very extensive survey of
Ref. \cite{bo07}.  If we calculated the theoretical $\dthree$ demanding
that the $j_z$ agree with the observed ground state spin, the values
would be some somewhat larger.  We note that in some cases the theoretical
ground state has mixed parity, so no spectroscopic identification is
possible.  For now, we take the
present disagreement on the spectroscopy identity of the levels 
as an indicator of the accuracy of the mean-field theory.
We anticipate extending the theory 
to restore angular momentum symmetry, and then a more complete
comparison of the spectroscopy could be made.

\section{Systematics}

To better see the overall trends in the pairing gaps, we 
show in Table I average values of the pairing gaps for each isotope
chain.
The range of measured gaps in the isotope chain and the experimental 
average is shown in the
second and third columns, respectively.  One sees that the
experimental gaps vary smoothly with the size of the nuclei, undergoing
a mild decrease as a function of $Z$.
If one takes the usual phenomenological parameterization of pairing gaps,
varying as
$A^{-1/2}$, one finds that the 
gaps for deformed nuclei are somewhat lower
than one would expect from the spherical $A$-systematics.  This in not surprising,
given that the pairing depends on the single-particle level density at the Fermi 
surface and the levels of deformed nuclei are more spread out .

The last three columns of the table show theoretical results for the
filling approximation and the two treatments of the $T$-odd fields.
The entries in the table are the differences of the average theoretical 
and experimental gaps, taking the same nuclei to make the averages.
For example, there are 4 measured gaps in the $Z=8$
chain with an average of $\dthree = 1.51$ MeV.  In the HFB theory with the
filling approximation the average gap of those triplets is 1.10 MeV for
an error of -0.41 MeV.

\begin{table}[htb] 
\caption{Average measured pairing gaps in selected isotope chains and
the errors in the corresponding quantities for the various treatments of
the time-odd fields.  
\label{systematics} }
\begin{tabular}{|cc|c|ccc|}
\colrule
Z   & N   & Exp.   &    Filling  & T-odd   &  T-odd \\
    & range&&& Pert. &   Full\\
\colrule
8   &   9-15  &       1.51 &    -0.41   &      -0.07  &   -0.36\\
50  &   53-83 &       1.18 &    +0.06   &      +0.22  &   +0.06\\
62  & 75-99     &       0.99 &    -0.29   &      -0.21  &   -0.30\\
82  &   97-131  &       0.98 &    +0.27   &      +0.37  &   +0.27\\
92  & 135-147   &       0.64 &    -0.19   &      -0.13  &   -0.21     \\
\colrule
\end{tabular} 
\end{table}

\begin{figure}[htb] 
\begin{center} 
\includegraphics[width=8cm]{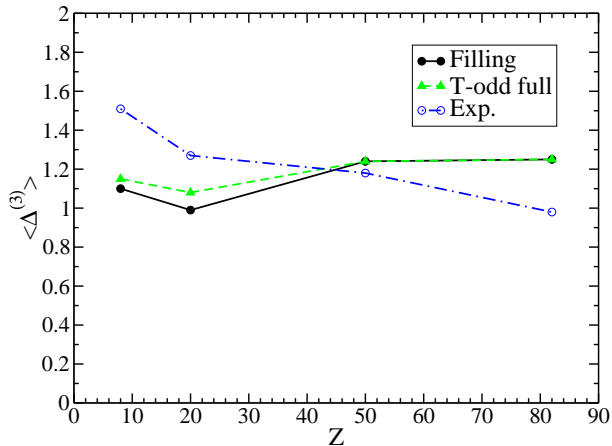} 
\caption{Average pairing gaps in the four spherical chains
$Z=8,20,50$ and 82, comparing theory and experiment.
\label{trend}} 
\end{center} 
\end{figure}

The errors for HFB+D1S theory in the filling approximation are shown
in the fourth column.   Both positive and negative errors are found,
which we attribute to a dependance on deformation and on size of the 
nucleus.  Namely, the theory predicts much smaller gaps in 
deformed nuclei than in the spherical ones.
Also, in the medium and
heavy nuclei the spherical nuclei
are somewhat overpredicted while the deformed nuclei are 
substantially underpredicted.
This highlights a deficiency of the HFB+D1S theory that cannot be
cured by simply adjusting the overall strength of the interaction
responsible for the pairing. Another deficiency of the theory 
can be seen by comparing the spherical chains, namely that
there is very little $Z$ dependence compared to experiment. This
be seen more clearly in Fig. \ref{trend}, which includes the
$Z=20$ chain along with the three other spherical chains in 
Table I .
The
lack of a significant $Z$ dependence affects in particular the
$Z=8$ chain, which is seriously underpredicted.  

The effect of including the $T$-odd field as a first-order
perturbation is shown in the fifth column.  The effect is to 
increase the pairing gaps, in qualitative agreement with the simple
model described in the appendix.  The increase is rather
uniform over the isotope chains in the table, and 
the above-discussed deficiencies remain.  

The full minimization of the HFB functional including the $T$-odd field 
gives gap errors shown in the last column of the Table.  One sees 
that the re-minimization
largely restores
the calculate gaps to the filling approximation values.  Thus,
the filling approximation seems to be accurate enough to assess
the HFB+D1S theory to the point of uncovering its systematic 
deficiencies.  
 
\section{Outlook}

With our implementation of the new technique \cite{ro11} for finding HFB 
minimum of odd-A nuclei we showed that it is a practical algorithm
for the interactions in common use.  Note that there is nothing in 
method that restricts the wave function to the one-quasiparticle space.  We
plan in the future to apply the method to two-quasiparticle wave functions,
starting with the ground states of odd-odd nuclei.  The energy splitting
of the states with different $K$ quantum numbers in strong deformed 
nuclei will provide a good test of the $T$-odd part of the interaction
\cite{bo76}.

Concerning the validity of the filling approximation, 
we found that $T$-odd fields of 
the Gogny D1S interaction have a small effect and their inclusion does not
noticeably improve the theoretical gaps.  The same conclusion can be
drawn for the $T$-odd field of a Skyrme interaction, from the 
HF-BCS study of gaps in Ref. \cite{po10} and 
HFB study of proton gaps in deformed nuclei in Ref. \cite{sc10}.

Concerning the performance of HFB on reproducing the experimental
pairing, we found that the overall strength of the pairing in the
D1S functional is close to optimal that it could reliably predict
gap quenching.  However, the mass dependence and the deformation
dependence seems to deviate from the observed phenomenology. 

This raises the question, what is missing in the theory that could be
significant for pairing gaps?  First of all, although the present $T$-odd 
effects are weak, the $T$-odd interaction should
be re-examined with a view to making better energy functionals.  In
particular, the mean-field contribution to pairing gaps 
could affect the overall $A$-dependence \cite{du01,fr07}.

There are a number of correlation effects that could affect the pairing
gaps.  The most obviously ones are those that restore broken symmetries
that may occur in the HFB wave functions.
For example, projection of good particle number has been 
shown to have a non-negligible effect on the performance of Skyrme
functionals \cite{an02,be09,mu11}. However, improvement comes mostly from 
nuclei having weak pairing condensates.  Restoration of  angular 
momentum symmetry can be very important, giving
rise to correlation energies of
the order of several MeV in deformed nuclei.  This is much 
larger than the ~0.1 MeV 
accuracy scale of pairing gap energies we would like to achieve.
Particularly critical to odd-even differences is the
presence of so-called Coriolis coupling effects in odd-$N$ systems
with small $j_z$ \cite[Ch. 4]{BM2}.  
Also, 
the energy gain by angular momentum projection is quite different in
spherical and deformed nuclei, so it could affect the gaps in transitional
nuclei.  We consider the problem of restoration of angular momentum
symmetry the most important computational issue to be address in future
work.  It is needed to make spectroscopic predictions, and it is 
needed to treat soft deformed nuclei on the same footing as the others.
Unfortunately, the computational effort to carry out angular momentum
restoration is heavy.

Another correlation is associated with the polarization of the 
nucleus by the valence nucleons.   The resulting 
induced pairing interaction has been calculated to give as strong
a contribution as the nucleon-nucleon interaction in the pairing
channel \cite{broglia},
Such induced
interactions are long-ranged and energy-dependent, and vary from nucleus
to nucleus depending on its structure.  It would not be surprising that
the effects were beyond the scope of the simple energy functionals 
in current use.

Lastly, there are correlation effects associated with the parts of
the Hamiltonian that are neglected in the HFB theory. In the theory
of the ground state, 4-quasiparticle excitations are neglected.
In the usual formulation
of the odd-$N$ theory in terms of quasiparticle excitations of the
even-$N$ wave functions, there is coupling to 3qp excitations that
give a significant contribution to the pairing gaps \cite{ku66}.  In our
formulation, the odd-$N$ wave function is an HFB local minimum and
therefore a quasiparticle vacuum.  The 2qp matrix elements of the
Hamiltonian vanish because of the minimization in the odd-$N$ space.
Thus only 4qp excitations need to be considered.  However, the quasiparticles 
energies can be negative in odd-$N$ systems, because the qp creation
operator can in effect unblock one of the orbitals.  This would give
smaller energy denominators in the second-order perturbative contribution 
for the $H^{40}$ term in the Hamiltonian.  
Besides reducing the pairing gaps, its contribution 
might have a different dependence on  $A$ and on deformation.

\section*{Acknowledgments}

This work was supported in part by the U.S. Department of Energy under Grant
DE-FG02-00ER41132, and by the National Science
Foundation under Grant PHY-0835543. 
The work of LMR was supported by MICINN (Spain) under  
grants Nos. FPA2009-08958, and FIS2009-07277, as well as by  
Consolider-Ingenio 2010 Programs CPAN CSD2007-00042 and MULTIDARK  
CSD2009-00064.  
\section*{Appendix}

The HFB energy contains a spurious
self-energy in the filling approximation for the unpaired particle 
in the wave function of an odd-A system. As a consequence, the filling 
approximation cannot be considered reliable for calculating quantities related to 
odd-even energy differences.  To see how this comes about, let us
consider a the $t_0$ term in the Skyrme interaction, i.e. the
simple $\delta$-function interaction $v(r_{12}) = t_0 \delta^3( r_{12})$.  
The density entering the Skyrme functional will be a matrix in 
spin and isospin, see e.g. \cite{be03}.  In the absence of spin-orbit
splitting, the density can be decomposed
into time-even and time-odd parts by dividing the orbitals into two
groups depending on the spin quantum number $s_z$.
Labeling the groups by $+$ and $-$, the $t_0$ contribution to the 
identical-particle energy
functional is 
\be
\langle v \rangle  = t_0 \int d^3 r \rho_+(r)\rho_-(r).
\ee
Next the 
densities are decomposed into time-even and time-odd densities
as $\rho_{e,o} = \rho_+ \pm \rho_-$.  Then
the interaction energy becomes
\be
\label{veo}
\langle v \rangle  = {t_0\over 4} \int d^3 r (\rho_e)^2 - {t_0\over 4} 
\int d^3 r (\rho_o)^2 .
\ee
For the paired ground states, the second term is nonzero when there
is an odd number of particles, but it is dropped in the 
filling approximation.  To assess the magnitude of the error, we examine the 
simplest cases: a) a state with one particle; b) the 
two-particle state in which a particle is put into the
time reversed orbital.  Let us call the $\rho_+$ density associated
with the one-particle state $\rho_1$.  Then 
\be
\rho_e = \rho_o = \rho_1  \,\,\,\,\,{\rm( 1~ particle)}.
\ee
Taking both T-even and T-odd terms in Eq. (\ref{veo}), the terms
cancel giving zero interaction of a particle with itself.  In the filling
approximation, the energy is 
\be
\langle v \rangle\rangle_{1,f} = {t_0\over 4} \int d^3 r (\rho_1)^2.
\ee
This may be compared with the two-particle interaction energy 
  $\langle v \rangle_2$ given by
Eq. (\ref{veo}) with 
\be
\rho_e = 2\rho_1; \,\,\,\,\rho_o = 0   \,\,\,\,{\rm (2~ particles)}.
\ee
From this it follows that the spurious self-energy is related to the
two-particle matrix element by
\be
\langle v \rangle_{1f}  = {1\over 4} \langle v \rangle_2.
\ee
This can only be small if the diagonal two-particle interaction matrix
elements are small.  

Of course in actual nuclei the spin-orbit field is very important,
contrary to what was assumed here.  Neverless, the model shows that
$T$-odd field can be significant, and are likely to be repulsive in
the perturbative limit for interactions that are attractive
in the filling approximation. We note further that the problem of 
spurious self-energies is particularly severe in theories that
do not have an accurate treatment of the exchange (Fock) interaction.
For example, in the Relativistic Mean Field theory \cite{af10},
the perturbative T-odd contribution is always negative, i.e. opposite
in sign to HFB.


\begin{thebibliography}{99}
\bibitem{ro11} L.M. Robledo and G.F. Bertsch, Phys. Rev. C 84 014312 (2011).
\bibitem{D1S} J.F.~Berger, M. Girod and D. Gogny, Comp. Phys. Comm.
{\bf 63} 365 (1991).
\bibitem{ba74} B.~Banerjee, P.~Ring, and H.J.~Mang, Nucl. Phys. {\bf A221}
564 (1974).
\bibitem{ma75} H.J.~Mang, Phys. Rep. {\bf 18} 325 (1975).
\bibitem{vi97} A.~Villafranca and J.L. Egido, Physics Letters B {\bf 408} 35 
(1997).
\bibitem{pe08}S.~Perez-Martin and L.M.~Robledo,
Phys. Rev. C, {\bf 78} 014304, (2008).
\bibitem{de80} J.~Decharge and D.~Gogny, Phys. Rev. C {\bf 21} 1568
(1980).
\bibitem{ro10a} R.~Rodriguez-Guzman, P.~Sarriguren, and L.~M. Robledo, 
Phys. Rev. C {\bf 82} 061302 (2010).
\bibitem{ro10b} R.~Rodriguez-Guzman, P.~Sarriguren, L.M. Robledo, and
S.~Perez-Martin, Phys. Lett. B {\bf 691} 202 (2010).
\bibitem{pa76} K.-Y.~Passler, Nucl. Phys. {\bf A257} 253 (1976).
\bibitem{be00} M.~Bender, K.~Rutz, P.~G.~Rienhard, and J.A.~Maruhn, 
Eur. Phys. J. A {\bf 8} 59 (2000).
\bibitem{be09} G.F.~Bertsch, C.A.~Bertulani, W.~Nazarewicz, N.~Schunck, and
M.V.~Stoitsov, Phys. Rev. C {\bf 79}, 034306 (2009).
\bibitem{po10} J.~Pototzky, J.~Erler,
P.-G.~Reinhard,
and V.O.~Nesterenko, Eur. Phys. J. A {\bf 46} 299 (2010).
\bibitem{sc10} N. Schunck, J. Dobaczewski, et al., Phys. Rev. C {\b 81}
024316 (2010).
\bibitem{ba06} E.~Baldini-Neto, B.V.~Carlson and D.~Hirata, J. Phys.
G {\b 32} 655 (2006).
\bibitem{ru99} K. Rutz, M. Bender, P-G~Reinhard, and J.A.~Maruhn, 
Phys. Lett. B468 1 (1999).
\bibitem{af10} A.V.~Afanasjev and H. Abusara, Phys. Rev. C {\bf 81}
014309 (2010).
\bibitem{du01} T.~Duguet, et al., Phys. Rev. C {\bf 65} 014311 (2001).
\bibitem{fr07} W.A. Friedman and G.F. Bertsch, Eur. Phys. J. A{\bf41} 109
(2009).
\bibitem{hfbaxial} L.M. Robledo, HFBAXIAL code (2002).
\bibitem{RS}  P.~Ring and P.~Schuck, ``The Nuclear Many-Body Problem'',
(Springer, Heidelberg, 1980).
\bibitem{CPC} L.M.~Robledo and G.F.~Bertsch, to be submitted to
Computer Physics Communications.
\bibitem{audi} G. Audi, A.H.~Wapstra, and C. Thibault, Nucl. Phys
{\bf A729} 337 (2003).
\bibitem{audi2011} Private communication by Georges Audi and Wang
Meng (2011); see {\tt http://amdc.in2p3.fr/masstables/Ame2011int/mass.mas114}
\bibitem{an02} M. Anguiano, J.L.~Egido, L.M.~Robledo, Phys. Lett. B{\bf 545} 62 (2002).
\bibitem{yu03} Y.~Yu and A.~Bulgac, Phys. Rev. Lett. {\bf 90} 222501 (2003).
\bibitem{bo07} L.~Bonneau, P.~Quentin, and P.~M\"oller, Phys. Rev. C
{\bf 76} 024320 (2007).
\bibitem{bo76} J.P.~Boisson, R.~Piepenbring and W.~Ogli, Physics Reports
{\bf 26} 99. (1976).
\bibitem{mu11} A. Mukherjee, Y. Alhassid, and G.F.~Bertsch, Phys. Rev. 
C {\bf 83} 014319 (2011).
\bibitem{BM2} A. Bohr and B. Mottelson, "Nuclear Structure, Vol. II" 
(Benjamin, New York, 1975), Chap. 4.
\bibitem{broglia} F. Barranco, et al., Eur. Phys. J. A {\bf 21} 57 (2004).
\bibitem{ku66}  T.~Kuo, E.~Baranger, and M.~Baranger, Nucl. Phys. {\bf
79} 513 (1966).
\bibitem{be03} M. Bender, P.-H. Heenen, and P.-G. Reinard, Rev. Mod.
Phys. 75 121 (2003).


\end{thebibliography}
\end{document}